\documentclass[letterpaper,twocolumn,10pt]{article}
\usepackage{usenix2019_v3}

\usepackage{verbatim}
\usepackage{booktabs}
\usepackage{tabularx}
\usepackage{mdwlist}
\usepackage{todonotes}
\usepackage{subcaption}
\usepackage{listings}
\usepackage{multirow}
\usepackage{makecell}
\usepackage{url}

\usepackage{ulem}
\usepackage[shortcuts]{extdash}
\usepackage{xcolor,colortbl}
\usepackage{graphicx}
\usepackage{balance}
\usepackage{paralist}

\usepackage{amssymb}
\usepackage{pifont}
\newcommand{\cmark}{\ding{51}}
\newcommand{\xmark}{\ding{55}}

\presetkeys{todonotes}{inline}{}

\newcommand \redbold[1]{\textcolor{red}{\textbf{#1}}}
\newcommand \bluebold[1]{\textcolor{blue}{\textbf{#1}}}

\usepackage[pdftex,
            pdfauthor={Seyed Ali Mirheidari, Sajjad Arshad, Kaan Onarlioglu, Bruno Crispo, Engin Kirda, William Robertson},
            pdftitle={Cached and Confused: Web Cache Deception in the Wild},
            pdfsubject={},
            pdfkeywords={cache deception, cache poisoning, CDN},
            pdfproducer={},
            pdfcreator={}]{hyperref}

\begin{document}

\title{Cached and Confused: Web Cache Deception in the Wild}

\author{
{\rm Seyed Ali Mirheidari}\\
University of Trento
\and
{\rm Sajjad Arshad}\thanks{Currently employed by Google.}\\
Northeastern University
\and
{\rm Kaan Onarlioglu}\\
Akamai Technologies
\and
{\rm Bruno Crispo}\\
University of Trento \&\\
KU Leuven
\and
{\rm Engin Kirda}\\
Northeastern University
\and
{\rm William Robertson}\\
Northeastern University
}

\maketitle

\begin{abstract}

Web cache deception (WCD) is an attack proposed in 2017, where an attacker
tricks a caching proxy into erroneously storing private information transmitted
over the Internet and subsequently gains unauthorized access to that cached
data. Due to the widespread use of web caches and, in particular, the use of
massive networks of caching proxies deployed by content distribution network
(CDN) providers as a critical component of the Internet, WCD puts a
substantial population of Internet users at risk.

We present the first large-scale study that quantifies the prevalence of WCD in
340~high-profile sites among the Alexa Top~5K. Our analysis reveals WCD
vulnerabilities that leak private user data as well as secret authentication and
authorization tokens that can be leveraged by an attacker to mount damaging web
application attacks. Furthermore, we explore WCD in a scientific framework as an
instance of the path confusion class of attacks, and demonstrate that variations
on the path confusion technique used make it possible to exploit sites that are
otherwise not impacted by the original attack. Our findings show that many
popular sites remain vulnerable two years after the public disclosure of WCD.

Our empirical experiments with popular CDN providers underline the fact that web
caches are not plug \& play technologies. In order to mitigate WCD, site
operators must adopt a holistic view of their web infrastructure and carefully
configure cache settings appropriate for their applications.
\end{abstract}

\section{Introduction}
\label{section:introduction}

Web caches
have become an
essential component of the Internet infrastructure with numerous use cases
such as reducing bandwidth costs in private enterprise networks and
accelerating content delivery over the World Wide Web. Today caching is
implemented at multiple stages of Internet communications, for instance in
popular web browsers~\cite{firefox,chromium}, at caching
proxies~\cite{varnish,squid}, and directly at origin web
servers~\cite{apache,nginx}.

In particular, Content Delivery Network~(CDN) providers heavily rely on
effective web content caching at their edge servers, which together comprise a 
massively-distributed Internet overlay network of caching reverse proxies.
Popular CDN providers advertise accelerated content delivery and high
availability via global coverage and deployments reaching hundreds of
thousands of servers~\cite{cloudflare_facts,akamai_facts}.  A recent
scientific measurement also estimates that more than 74\% of the Alexa Top
1K are served by CDN providers, indicating that CDNs and more generally
web caching play a central role in the Internet~\cite{guo2018}.

While there exist technologies that enable limited caching of
dynamically-generated pages, web caching primarily targets static, publicly
accessible content. In other words, web caches store static content that is
costly to deliver due to an object's size or distance.  Importantly, these objects
\textit{must not} contain private or otherwise sensitive information, as application-level
access control is not enforced at cache servers.  Good
candidates for caching include frequently accessed images, software and document
downloads, streaming media, style sheets, and large static HTML and JavaScript
files.

In 2017, Gil presented a novel attack called \textit{web cache
deception~(WCD)} that can trick a web cache into incorrectly storing
sensitive content, and consequently give an attacker unauthorized access
to that content~\cite{wcd_blog,wcd_paper}. Gil demonstrated the issue with
a real-life attack scenario targeting a high profile site, PayPal, and
showed that WCD can successfully leak details of a private payment
account. Consequently, WCD garnered significant media attention, and
prompted responses from major web cache and CDN
providers~\cite{cloudflare_wcd1,cloudflare_wcd2,akamai_wcd,symantec_wcd,mnot_wcd,bleepingcomputer_wcd}.

At its core, WCD results from \textit{path confusion} between an origin server
and a web cache. In other words, different interpretations of a requested URL at
these two points lead to a disagreement on the cacheability of a given object.
This disagreement can then be exploited to trick the web cache into storing
non-cacheable objects. WCD does not imply that these individual components---the
origin server and web cache---are incorrectly configured per se. Instead, their
hazardous interactions as a system lead to the vulnerability. As a result,
detecting and correcting vulnerable systems is a cumbersome task, and may
require careful inspection of the entire caching architecture. Combined with the
aforementioned pervasiveness and critical role of web caches in the Internet
infrastructure, WCD has become a severely damaging issue.

In this paper, we first present a large-scale measurement and analysis of WCD
over 295~sites in the Alexa Top~5K. We present a repeatable and automated
methodology to discover vulnerable sites over the Internet, and a detailed
analysis of our findings to characterize the extent of the problem. Our results
show that many high-profile sites that handle sensitive and private data are
impacted by WCD and are vulnerable to practical attacks. We then discuss
additional path confusion methods that can maximize the damage potential of WCD,
and demonstrate their impact in a follow-up experiment over an extended data set
of 340~sites.

To the best of our knowledge, this is the first in-depth investigation of WCD in
a scientific framework and at this scale. In addition, the scope of our
investigation goes beyond private data leakage to provide novel insights into
the severity of WCD. We demonstrate how WCD can be exploited to steal other
types of sensitive data including security tokens, explain advanced attack
techniques that elevate WCD vulnerabilities to injection vectors, and quantify
our findings through further analysis of collected data.

Finally, we perform an empirical analysis of popular CDN providers, documenting
their default caching settings and customization mechanisms. Our findings
underline the fact that WCD is a \textit{system safety} problem. Site operators
must adopt a holistic view of their infrastructure, and carefully configure web
caches taking into consideration their complex interactions with origin servers.

To summarize, we make the following contributions:

\begin{itemize}

\item We propose a novel methodology to detect sites impacted by WCD at
  scale. Unlike existing WCD scan tools that are designed for site
  administrators to test their own properties in a controlled environment, our
  methodology is designed to automatically detect WCD in the wild.

\item We present findings that quantify the prevalence of WCD in 295~sites among
  the Alexa Top 5K, and provide a detailed breakdown of leaked information
  types. Our analysis also covers security tokens that can be stolen via WCD as
  well as novel security implications of the attack, all areas left unexplored
  by existing WCD literature.

\item We conduct a follow-up measurement over 340~sites among the Alexa Top~5K
  that show variations on the path confusion technique make it possible to
  successfully exploit sites that are not impacted by the original attack.

\item We analyze the default settings of popular CDN providers and document
  their distinct caching behavior, highlighting that mitigating WCD necessitates
  a comprehensive examination of a website's infrastructure.

\end{itemize}

\textbf{Ethical Considerations.} We have designed our measurement methodology to
minimize the impact on scanned sites, and limit the inconvenience we impose on
site operators. Similarly, we have followed responsible disclosure principles to
notify the impacted parties, and limited the information we share in this paper
to minimize the risk of any inadvertent damage to them or their end-users. We
discuss details of the ethical considerations pertaining to this work in
Section~\ref{section:ethics}.

\section{Background \& Related Work}
\label{background}

In this section, we present an overview of how web cache deception~(WCD)
attacks work and discuss related concepts and technologies such as web
caches, path confusion, and existing WCD scanners. As of this writing, the
academic literature has not yet directly covered WCD.  Nevertheless, in
this section we summarize previous publications pertaining to other
security issues around web caches and CDNs.

\subsection{Web Caches}
\label{subsec:cache}

Repeatedly transferring heavily used and large web objects over the Internet is
a costly process for both web servers and their end-users. Multiple round-trips
between a client and server over long distances, especially in the face of
common technical issues with the Internet infrastructure and routing
problems, can lead to increased network latency and result in web
applications being perceived as unresponsive. Likewise, routinely accessed
resources put a heavy load on web servers, wasting valuable computational
cycles and network bandwidth. The Internet community has long been aware
of these problems, and deeply explored caching strategies and technologies
as an effective solution.

Today web caches are ubiquitous, and are used at various---and often
multiple---steps of Internet communications. For instance, client
applications such as web browsers implement their own \textit{private}
cache for a single user.  Otherwise, web caches deployed together with
a web server, or as a man-in-the-middle proxy on the communication path
implement a \textit{shared} cache designed to store and serve objects
frequently accessed by multiple users.  In all cases, a cache hit
eliminates the need to request the object from the origin server,
improving performance for both the client and server.

In particular, web caches are a key component of Content Delivery Networks~(CDN)
that provide web performance and availability services to their users. By
deploying massively-distributed networks of shared caching proxies (also called
\textit{edge servers}) around the globe, CDNs aim to serve as many requests as
possible from their caches deployed closest to clients, offloading the origin
servers in the process. As a result of multiple popular CDN providers that cover
different market segments ranging from simple personal sites to large
enterprises, web caches have become a central component of the Internet
infrastructure. A recent study by Guo et al.~estimates that 74\% of the Alexa Top 1K
make use of CDNs~\cite{guo2018}.

The most common targets for caching are static but frequently accessed
resources. These include static HTML pages, scripts and style sheets,
images and other media files, and large document and software downloads.
Due to the shared nature of most web caches, objects containing dynamic,
personalized, private, or otherwise sensitive content are not suitable for
caching. We point out that there exist technologies such as Edge Side
Includes~\cite{esi} that allow caching proxies to assemble responses from
a cached static part and a freshly-retrieved dynamic part, and the
research community has also explored caching strategies for dynamic
content. That being said, caching of non-static objects is not common, and
is not relevant to WCD attacks.  Therefore, it will not be discussed further
in this paper.

The HTTP/1.1 specification defines \texttt{Cache-Control} headers that can
be included in a server's response to signal to all web caches on the
communication path how to process the transferred
objects~\cite{ietfcache}. For example, the header \texttt{``Cache-
Control: no-store''} indicates that the response should not be stored.
While the specification states that web caches \textit{MUST} respect these
headers, web cache technologies and CDN providers offer configuration
options for their users to ignore and override header instructions.
Indeed, a common and easy configuration approach is to create simple
caching rules based on resource paths and file names, for instance,
instructing the web cache to store all files with extensions such as
\verb+jpg+, \verb+ico+, \verb+css+, or \verb+js+~\cite{akamai_cacheoverride,cloudflare_cacheoverride}.

\subsection{Path Confusion}
\label{subsec:pathconfusion}

Traditionally, URLs referenced web resources by directly mapping these to
a web server's filesystem structure, followed by a list of query
parameters. For instance, \texttt{example.com/home/index.html?lang=en}
would correspond to the file \texttt{home/index.html} at that web server's
document root directory, and \texttt{lang=en} represents a parameter
indicating the preferred language.

However, as web applications grew in size and complexity, web servers
introduced sophisticated URL rewriting mechanisms to implement
advanced application routing structures as well as to improve usability
and accessibility. In other words, web servers parse, process, and
interpret URLs in ways that are not clearly reflected in the
externally-visible representation of the URL string.  Consequently, the
rest of the communication endpoints and man-in-the-middle entities may
remain oblivious to this additional layer of abstraction between the
resource filesystem path and its URL, and process the URL in an
unexpected---and potentially unsafe---manner. This is called \textit{path
confusion}.

The widespread use of \textit{clean URLs} (also known as \textit{RESTful
URLs}) help illustrate this disconnect and the subsequent issues
resulting from different interpretations of a URL. Clean URL schemes use
structures that abstract away from a web server's internal organization of
resources, and instead provide a more readable API-oriented
representation.  For example, a given web service may choose to structure
the URL \texttt{example.com/index.php?p1=v1\&p2=v2} as
\texttt{example.com/index/v1/v2} in clean URL representation. Now,
consider the case where a user accesses the same web service using the URL
\texttt{example.com/index/img/pic.jpg}. The user and all technologies in
the communication path (e.g., the web browser, caches, proxies, web
application firewalls) are likely to misinterpret this request, expect an
image file in return, and treat the HTTP response accordingly (e.g., web
caches may choose to store the response payload). However, in reality, the
web service will internally map this URL to
\texttt{example.com/index.php?p1=img\&p2=pic.jpg}, and return the contents
of \texttt{index.php} with an HTTP 200 status code. Note that even when
\texttt{img/pic.jpg} is an arbitrary resource that does not exist on the
web server, the HTTP 200 status code will falsely indicate that the
request was successfully handled as intended.

Web application attacks that involve malicious payload injection, such as
cross-site scripting, are well-understood and studied by both academics
and the general security community. Unfortunately, the security
implications of path confusion have started to garner attention only
recently, and academic literature on the subject is sparse.

One notable class of attacks based on path confusion is \textit{Relative
Path Overwrite (RPO)}, first presented by Gareth Heyes in 2014~\cite{rpo}.
RPO targets sites that utilize relative paths for security-sensitive
resource inclusions such as style sheets and scripts. The attack is made
possible by maliciously-crafted URLs that are still interpreted in the
same way their benign counterparts are by web servers, but when used as
the base URL causes a web browser to expand relative paths incorrectly.
This results in attacker-controlled same-origin inclusions. Other
researchers have since proposed variations on more advanced applications
of RPO, which can elevate this attack vector into numerous other 
vulnerabilities~\cite{rpo1,rpo2,rpo3,rpo4}. Recently, Arshad et
al.~conducted a large-scale measurement study of RPO in the wild and
reported that 9\% of the Alexa Top 1M are vulnerable, and that more than one
third of these are exploitable~\cite{arshad2018}.

Other related work include more general techniques for exploiting URL parser 
behavior. For instance, Orange Tsai presented a series of exploitation
techniques that take advantage of the quirks of built-in URL parsers in popular
programming languages and web frameworks~\cite{tsai1,tsai2}. While Tsai's
discussion mainly focuses on Server-Side Request Forgery, these techniques are
essentially instances of path confusion and can be utilized in many attacks in
the category.

Our focus in this paper is web cache deception, the most recently
discovered major security issue that is enabled by an attacker exploiting
a path confusion vulnerability. To the best of our knowledge, this paper
is the first academic exploration of WCD in the literature, and also
constitutes the first large-scale analysis of its spread and severity.

\begin{figure*}
    \centering
    \includegraphics[width=0.80\textwidth]{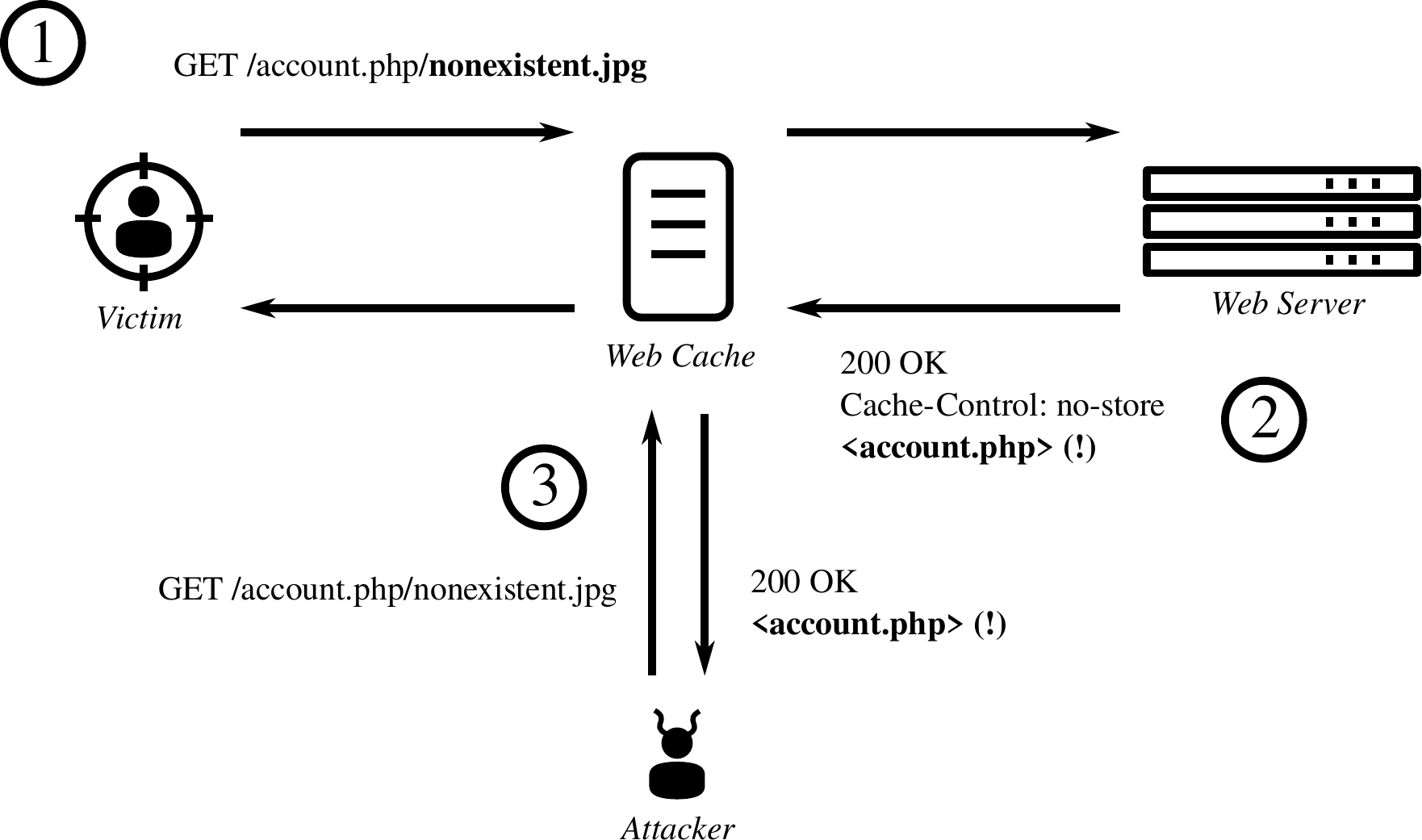}
    \caption{An illustrated example of web cache deception. Path confusion
      between a web cache and a web server leads to unexpected caching of the
      victim's private account details. The attacker can then issue a request
      resulting in a cache hit, gaining unauthorized access to cached private
      information.}
    \label{fig:wcd-background}
\end{figure*}

\subsection{Web Cache Deception}
\label{subsec:wcd}

WCD is a recently-discovered manifestation of path confusion that an attacker
can exploit to break the confidentiality properties of a web application. This
may result in unauthorized disclosure of private data belonging to end-users of
the target application, or give the attacker access to sensitive security tokens
(e.g., CSRF tokens) that could be used to facilitate further web application
attacks by compromising authentication and authorization mechanisms. Gil
proposed WCD in 2017, and demonstrated its impact with a practical attack
against a major online payment provider, PayPal~\cite{wcd_blog,wcd_paper}.

In order to exploit a WCD vulnerability, the attacker crafts a URL that
satisfies two properties:
\begin{enumerate}

\item The URL must be interpreted by the web server as a request for a
non-cacheable page with private information, and it should trigger a successful
response.

\item The same URL must be interpreted by an intermediate web cache as
a request for a static object matching the caching rules in effect.

\end{enumerate}

Next, the attacker uses social engineering channels to lure a victim into
visiting this URL, which would result in the incorrect caching of the victim's
private information. The attacker would then repeat the request and gain access
to the cached contents. Figure~\ref{fig:wcd-background} illustrates these
interactions.

\begin{inparaenum}[\itshape Step 1\upshape]

In \item, the attacker tricks the victim into visiting a URL that requests
\texttt{/account.php/nonexistent.jpg}. At a first glance this appears to
reference an image file, but in fact does not point to a valid resource on the
server.

In \item, the request reaches the web server and is processed. The server in
this example applies rewrite rules to discard the non-existent part of the
requested object, a common default behavior for popular web servers and
application frameworks. As a result, the server sends back a success response,
but actually includes the contents of \texttt{account.php} in the body, which
contains private details of the victim's account. Unaware of the URL mapping
that happened at the server, the web cache stores the response, interpreting it
as a static image.

Finally, in \item, the attacker visits the same URL which results in a cache hit
and grants him unauthorized access to the victim's cached account information.

\end{inparaenum}

Using references to non-existent cacheable file names that are interpreted
as path parameters is an easy and effective path confusion technique to
mount a WCD attack, and is the original attack vector proposed by Gil.
However, we discuss novel and more advanced path confusion strategies in
Section~\ref{sec:advanced_path_confusion}. Also note that the presence of
a \texttt{Cache-Control: no-store} header value has no impact in our
example, as it is common practice to enable caching rules on proxy
services that simply ignore header instructions and implement aggressive
rules based on path and file extension patterns (see
Section~\ref{subsec:cache}).

WCD garnered significant media attention due to its security implications and
high damage potential. Major web cache technology and CDN providers also
responded, and some published configuration hardening guidelines for their
customers~\cite{cloudflare_wcd1,akamai_wcd,symantec_wcd}. More recently,
Cloudflare announced options for new checks on HTTP response content types to
mitigate the attack~\cite{cloudflare_wcd2}.

Researchers have also published tools to scan for and detect WCD, for instance,
as an extension to the Burp Suite scanner or as stand-alone
tools~\cite{wcd_scan1,wcd_scan2}. We note that these tools are oriented towards
penetration testing, and are designed to perform targeted scans on web
properties directly under the control of the tester. That is, by design, they
operate under certain pre-conditions, perform information disclosure tests via
simple similarity and edit distance checks, and otherwise require manual
supervision and interpretation of the results. This is orthogonal to the
methodology and findings we present in this paper. Our experiment is, instead,
designed to discover WCD vulnerabilities at scale in the wild, and does not rely
on page similarity metrics that would result in an overwhelming number of false
positives in an uncontrolled test environment.

\subsection{Other Related Work}

Caching mechanisms in many Internet technologies (e.g., ARP, DNS) have been
targeted by \textit{cache poisoning} attacks, which involve an attacker storing
a malicious payload in a cache later to be served to victims. For example, James
Kettle recently presented practical cache poisoning attacks against caching
proxies~\cite{cache_poisoning,httpdesync}. Likewise, Nguyen et al.~demonstrated
that negative caching (i.e., caching of 4xx or 5xx error responses) can be
combined with cache poisoning to launch denial-of-service
attacks~\cite{nguyen2019}. Although the primary goal of a cache poisoning attack
is malicious payload injection and not private data disclosure, these attacks
nevertheless manipulate web caches using mechanisms similar to web cache
deception. Hence, these two classes of attacks are closely related.

More generally, the complex ecosystem of CDNs and their critical position as
massively-distributed networks of caching reverse proxies have been studied in
various security contexts~\cite{stocker2017,guo2018}. For example, researchers
have explored ways to use CDNs to bypass Internet
censorship~\cite{Holowczak2015,Zolfaghari2016,Fifield2015}, exploit or weaponize
CDN resources to mount denial-of-service attacks~\cite{Triukose2009,Chen2016},
and exploit vectors to reveal origin server addresses behind
proxies~\cite{Vissers2015,Jin2018}. On the defense front, researchers have
proposed techniques to ensure the integrity of data delivered over untrusted
CDNs and other proxy services~\cite{Lesni2003,Levy2016,Michalakis2007}.  This
research is orthogonal to WCD, and is not directly relevant to our results.

\section{Methodology}
\label{sec:methodology}

\begin{figure*}[t]
  \center
    \includegraphics[width=0.80\textwidth]{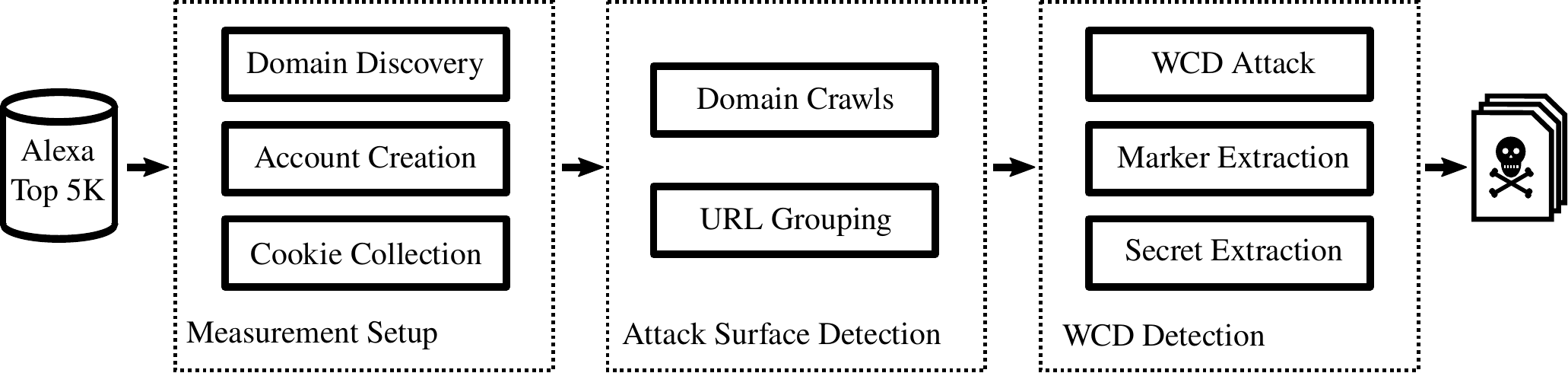}
    \caption{A high-level overview of our WCD measurement methodology.}
    \label{fig:methodology}
\end{figure*}

We present our measurement methodology in three stages:
\begin{inparaenum}[\itshape (1)\upshape]
\item measurement setup,
\item attack surface detection, and
\item WCD detection.
\end{inparaenum}
We illustrate this process in Figure~\ref{fig:methodology}. We implemented the
tools that perform the described tasks using a combination of Google Chrome and
Python's Requests library~\cite{python_requests} for web interactions, and
Selenium~\cite{selenium} and Google Remote Debugging Protocol~\cite{google_rdp}
for automation.

\subsection{Stage 1: Measurement Setup}

WCD attacks are only meaningful when a vulnerable site manages private end-user
information and allows performing sensitive operations on this
data. Consequently, sites that provide authentication mechanisms are prime
targets for attacks, and thus also for our measurements. The first stage of our
methodology identifies such sites and creates test accounts on them.\footnote{In
  the first measurement study we present in Section~\ref{sec:analysis}, we scoped
  our investigation to sites that support Google OAuth~\cite{google_oauth} for
  authentication due to its widespread use. This was a design choice made to
  automate a significant chunk of the initial account setup workload, a
  necessity for a large-scale experiment. In our follow-up experiment later
  described in Section~\ref{sec:advanced_path_confusion} we supplemented this data
  set with an additional 45 sites that do not use Google OAuth. We discuss these
  considerations in their corresponding sections.}

\paragraph{Domain Discovery.}

This stage begins by visiting the sites in an initial measurement
\textit{seed pool} (e.g., the Alexa Top~$n$ domains). We then increase site
coverage by performing sub-domain discovery using open-source intelligence
tools~\cite{amass,sublister,aquatone}. We add these newly-discovered sub-domains
of the primary sites (filtered for those that respond to HTTP(s) requests) to
the seed pool.

\paragraph{Account Creation.}

Next, we create two test accounts on each site: one for a \textit{victim},
and the other for an \textit{attacker}. We populate each account with unique
dummy values. Next, we manually explore each victim account to discover data
fields that should be considered private information (e.g., name, email,
address, payment account details, security questions and responses) or
user-created content (e.g., comments, posts, internal messages). We populate
these fields with predefined \textit{markers} that can later be searched for in
cached responses to detect a successful WCD attack. On the other hand, no data
entry is necessary for attacker accounts.

\paragraph{Cookie Collection.}

Once successfully logged into the sites in our seed pool, crawlers collect
two sets of cookies for all victim and attacker accounts. These are saved in a
cookie jar to be reused in subsequent steps of the measurement. Note that we
have numerous measures to ensure our crawlers remain authenticated during our
experiments. Our crawlers periodically re-authenticate, taking into account
cookie expiration timestamps. In addition, the crawlers use regular expressions
and blacklists to avoid common logout links on visited pages.

\subsection{Stage 2: Attack Surface Detection}

\paragraph{Domain Crawls.}

In the second stage, our goal is to map from domains in the seed pool to a set
of pages (i.e., complete URLs) that will later be tested for WCD
vulnerabilities. To this end, we run a recursive crawler on each domain in the
seed pool to record links to pages on that site.

\begin{table}[t]
  \footnotesize
    \caption{Sample URL grouping for attack surface discovery.}
    \label{tab:grouped_webpages}
    \centering
    \begin{tabular}{ll}
    \toprule
    \multicolumn{1}{c}{\textbf{Group By}} &
    \multicolumn{1}{c}{\textbf{URL}} \\
    \midrule
    \multirow{2}{*}{Query Parameter} & http://example.com/?lang=\textbf{en} \\
    & http://example.com/?lang=\textbf{fr} \\
    \midrule
    \multirow{2}{*}{Path Parameter} & http://example.com/\textbf{028} \\
    & http://example.com/\textbf{142} \\
    \bottomrule
    \end{tabular}
\end{table}

\paragraph{URL Grouping.}

Many modern web applications customize pages based on query string or URL path
parameters. These pages have similar structures and are likely to expose similar
attack surfaces. Ideally, we would group them together and select only one
random instance as a representative URL to test for WCD in subsequent steps.

Since performing a detailed content analysis is a costly process that could
generate an unreasonable amount of load on the crawled site, our URL grouping
strategy instead focuses on the structure of URLs, and approximates page
similarity without downloading each page for analysis. Specifically, we convert
the discovered URLs into an abstract representation by grouping those URLs by
query string parameter names or by numerical path parameters. We select one
random instance and filter out the rest. Table~\ref{tab:grouped_webpages}
illustrates this process.

This filtering of URLs significantly accelerates the measurements, and also
avoids overconsumption of the target site's resources with redundant scans in
Stage~3. We stop attack surface detection crawls after collecting 500~unique
pages per domain for similar reasons.

\subsection{Stage 3: WCD Detection}
\label{tab:methodology:detection}

In this final stage, we launch a WCD attack against every URL discovered in
Stage~2, and analyze the response to determine whether a WCD vulnerability was
successfully exploited.

\paragraph{WCD Attack.}

The attack we mount directly follows the scenario previously described in
Section~\ref{subsec:wcd} and illustrated in Figure~\ref{fig:wcd-background}.
For each URL:

\begin{enumerate}

\item We craft an attack URL that references a non-existent static resource. In
  particular, we append to the original page
  \texttt{``/<random>.css''}\footnote{Our choice to use a style sheet in our
    payload is motivated by the fact that style sheets are essential components
    of most modern sites, and also prime choices for caching. They are also a
    robust choice for our tests. For instance, many CDN providers offer
    solutions to dynamically resize image files on the CDN edge depending on the
    viewport of a requesting client device. Style sheets are unlikely to be
    manipulated in such ways.}. We use a random string as the file name in order
  to prevent ordinary end-users of the site from coincidentally requesting
  the same resource.

\item We initiate a request to this attack URL from the \textit{victim} account
and record the response.

\item We issue the same request from the \textit{attacker} account, and save the
response for comparison.

\item Finally, we repeat the attack as an \textit{unauthenticated user}
by omitting any session identifiers saved in the attacker cookie jar. We
later analyze the response to this step to ascertain whether attackers
without authentication credentials (e.g., when the site does not offer
open or free sign ups) can also exploit WCD vulnerabilities.

\end{enumerate}

\paragraph{Marker Extraction.}

Once the attack scenario described above is executed, we first check for
private information disclosure by searching the attacker response for the
\textit{markers} that were entered into victim accounts in Stage~1. If victim
markers are present in URLs requested by an attacker account, the attacker must
have received the victim's incorrectly cached content and, therefore, the
target URL contains an exploitable WCD vulnerability.  Because these markers
carry relatively high entropy, it is probabilistically highly unlikely that
this methodology will produce false positives.

\paragraph{Secret Extraction.}

We scan the attacker response for the disclosure of secret tokens frequently
used as part of web application security mechanisms. These checks include common
secrets (e.g., CSRF tokens, session identifiers) as well as any other
application-specific authentication and authorization tokens (e.g., API
credentials). We also check for session-dependent resources such as
dynamically-generated JavaScript, which may have private information and secrets
embedded in them (e.g., as explored by Lekies et al.~\cite{lekies2015}).

\begin{inparaenum}[\itshape (1)\upshape]%
In order to extract candidates for leaked secrets, we scan attacker responses
for name \& value pairs, where either \item the name contains one of our
keywords (e.g., \texttt{csrf}, \texttt{xsrf}, \texttt{token}, \texttt{state},
\texttt{client\_id}), or \item the value has a random component. We check for
these name \& value pairs in hidden HTML form elements, query strings extracted
from HTML anchor elements, and inline JavaScript variables and constants.
\end{inparaenum}
Similarly, we extract random file names referenced in HTML script elements. We
perform all tests for randomness by first removing dictionary words from the
target string (i.e., using a list of 10,000 common English
words~\cite{10k_common_words}), and then computing Shannon entropy over the
remaining part.

Note that unlike our checks for private information leaks, this process can
result in false positives. Therefore, we perform this secret extraction process
only when the victim and attacker responses are identical (a strong indicator of
caching), or otherwise when we can readily confirm a WCD vulnerability by
searching for the private information markers. In addition, we later manually
verify all candidate secrets extracted in this step.

\subsection{Verification and Limitations}

Researchers have repeatedly reported that large-scale Internet measurements,
especially those that use automated crawlers, are prone to being blocked or
served fake content by security solutions designed to block malicious bots and
content scrapers~\cite{kaan_bots,Wang2011}. In order to minimize this risk
during our measurement, we used a real browser (i.e., Google Chrome) for most
steps in our methodology. For other interactions, we set a valid Chrome
user-agent string. We avoided generating excessive amounts of traffic and
limited our crawls as described above in order to avoid triggering rate-limiting
alerts, in addition to ethical motivations. After performing our measurements,
we manually verified \textit{all} positive findings and confirmed the discovered
vulnerabilities.

Note that this paper has several important limitations, and the findings should
be considered a potentially loose lower bound on the incidence of WCD
vulnerabilities in the wild.  For example, as described in
Section~\ref{sec:analysis}, our seed pool is biased toward sites that support
Google OAuth, which was a necessary compromise to automate our methodology and
render a large-scale measurement feasible. Even under this constraint, creating
accounts on some sites required entering and verifying sensitive information
such as credit card or US social security numbers which led to their exclusion
from our study.

Furthermore, decisions such as grouping URLs based on their structure without
analyzing page content, and limiting site crawls to 500 pages may have caused us
to miss additional instances of vulnerabilities. Similarly, even though we
manually filtered out false positives during our secret token extraction process
and verified all findings, we do not have a scalable way of detecting false
\textit{negatives}. We believe that these trade-offs were worthwhile given the
overall security benefits of and lessons learned from our work. We emphasize
that the results in this paper represent a lower bound.

\subsection{Ethical Considerations}
\label{section:ethics}

Here, we explain in detail important ethical considerations pertaining to this
work and the results we present.

\paragraph{Performance Considerations.}

We designed our methodology to minimize the performance impact on scanned sites
and inconvenience imposed on their operators. We did not perform repeated or
excessive automated scans of the targeted sites, and ensured that our
measurements did not generate unreasonable amounts of traffic. We used only
passive techniques for sub-domain enumeration and avoided abusing external
resources or the target site's DNS infrastructure.

Similarly, our stored modifications to crawled web applications only
involved creating two test accounts and filling out editable fields with
markers that we later used for data leakage detection. We believe this
will have no material impact on site operators, especially in the
presence of common threats such as malicious bots and credential stuffing
tools that generate far more excessive junk traffic and data.

\paragraph{Security Considerations.}

Our methodology entirely avoids jeopardizing the security of crawled
sites or their end-users. In this work, we never injected or stored any
malicious payload to target sites, to web caches on the communication
path, or otherwise maliciously tampered with any technology involved in
the process. Likewise, the experiments we performed all incorporated
randomized strings as the non-existent parts of URLs, thereby preventing
unsuspecting end-users from accidentally accessing our cached data and
receiving unexpected responses.

Note that this path randomization measure was used to prevent inconveniencing or
confusing end-users; since we never exploited WCD to leak real personal data
from a web application or stored a malicious payload, our work never posed a
security risk to end-users.

Our experiments did not take into account robots.txt files. This was a
risk-based decision we consciously made, and we believe that ignoring exclusion
directives had no negative impact on the privacy of these sites'
visitors. Robots.txt is not a security or privacy mechanism, but is intended to
signal to data aggregators and search engines what content to index -- including
a directive to exclude privacy sensitive pages would actually be a misuse of
this technology. This is not relevant to our experiments, as we only collect
content for our analysis, and we do not index or otherwise publicly present site
content.

\paragraph{Responsible Disclosure.}

In this paper, we present a detailed breakdown of our measurement findings
and results of our analysis, but we refrain from explicitly naming the
impacted sites. Even though our methodology only utilized harmless
techniques for WCD detection, the findings point at real-world
vulnerabilities that could be severely damaging if publicly disclosed
before remediation.

We sent notification emails to publicly listed security contacts of all
impacted parties promptly after our discovery. In the notification letters
we provided an explanation of the vulnerability with links to online
resources and listed the vulnerable domain names under ownership of the
contacted party. We informed them of our intention to publicly publish
these results, noted that they will not be named, and advised that they
remediate the issue as adversaries can easily repeat our experiment and
compromise their sites. We also explicitly stated that we did not seek
or accept bug bounties for these notifications.

We sent the notification letters prior to submitting this work for review,
therefore giving the impacted parties reasonably early notice. As of this
writing, 12~of the impacted sites have implemented mitigations.

\paragraph{Repeatability.}

One of the authors of this paper is affiliated with a major CDN provider at the
time of writing. However, the work and results we present in this paper do not
use any internal or proprietary company information, or any such information
pertaining to the company's customers. We conducted this work using only
publicly available data sources and tools. Our methodology is repeatable by
other researchers without access to any CDN provider internals.

\section{Web Cache Deception Measurement Study}
\label{sec:analysis}

We conducted two measurement studies to characterize web cache deception~(WCD)
vulnerabilities on the Internet. In this first study we present in this section,
the research questions we specifically aim to answer are:

\begin{enumerate*}

\item[\textbf{(Q1)}] What is the prevalence of WCD vulnerabilities on popular,
highly-trafficked domains? (\S\ref{sec:analysis:summary})

\item[\textbf{(Q2)}] Do WCD vulnerabilities expose PII and, if so, what kinds?
(\S\ref{sec:analysis:vulns})

\item[\textbf{(Q3)}] Can WCD vulnerabilities be used to defeat defenses against
web application attacks? (\S\ref{sec:analysis:vulns})

\item[\textbf{(Q4)}] Can WCD vulnerabilities be exploited by unauthenticated
users? (\S\ref{sec:analysis:vulns})

\end{enumerate*}

In the following, we describe the data we collected to carry out the study. We
discuss the results of the measurement, and then consider implications for PII
and important web security defenses. Finally, we summarize the conclusions we
draw from the study. In Section~\ref{sec:advanced_path_confusion}, we will present
a follow-up experiment focusing on advanced path confusion techniques.

\subsection{Data Collection}
\label{sec:analysis:data}

\begin{table}[t]
    \caption{Summary of crawling statistics.}
    \label{tab:analysis:dataset}
    \centering
	\footnotesize
    \begin{tabular}{lrr}
    \toprule

    &
    \multicolumn{1}{c}{\textbf{Crawled}} &
    \multicolumn{1}{c}{\textbf{Vulnerable}}
    \\

    \midrule

    Pages & 1,470,410 & 17,293 (1.2\%) \\
    Domains & 124,596 & 93 (0.1\%) \\
    Sites & 295 & 16 (5.4\%) \\

    \bottomrule
    \end{tabular}
\end{table}

We developed a custom web crawler to collect the data used in this measurement.
The crawler ran from April~20-27, 2018 as a Kubernetes pod that was allocated 16
Intel Xeon 2.4~GHz CPUs and 32~GiB of RAM. Following the methodology described
in Section~\ref{sec:methodology}, we configured the crawler to identify vulnerable
sites from the Alexa Top 5K at the time of the experiment. In order to scalably
create test accounts, we filtered this initial measurement seed pool for sites
that provide an option for user authentication via Google OAuth. This filtering
procedure narrowed the set of sites considered in this measurement to~295.
Table~\ref{tab:analysis:dataset} shows a summary of our crawling statistics.

\subsection{Measurement Overview}
\label{sec:analysis:summary}

\paragraph{Alexa Ranking.}

\begin{figure}
    \centering
    \includegraphics[width=1\linewidth]{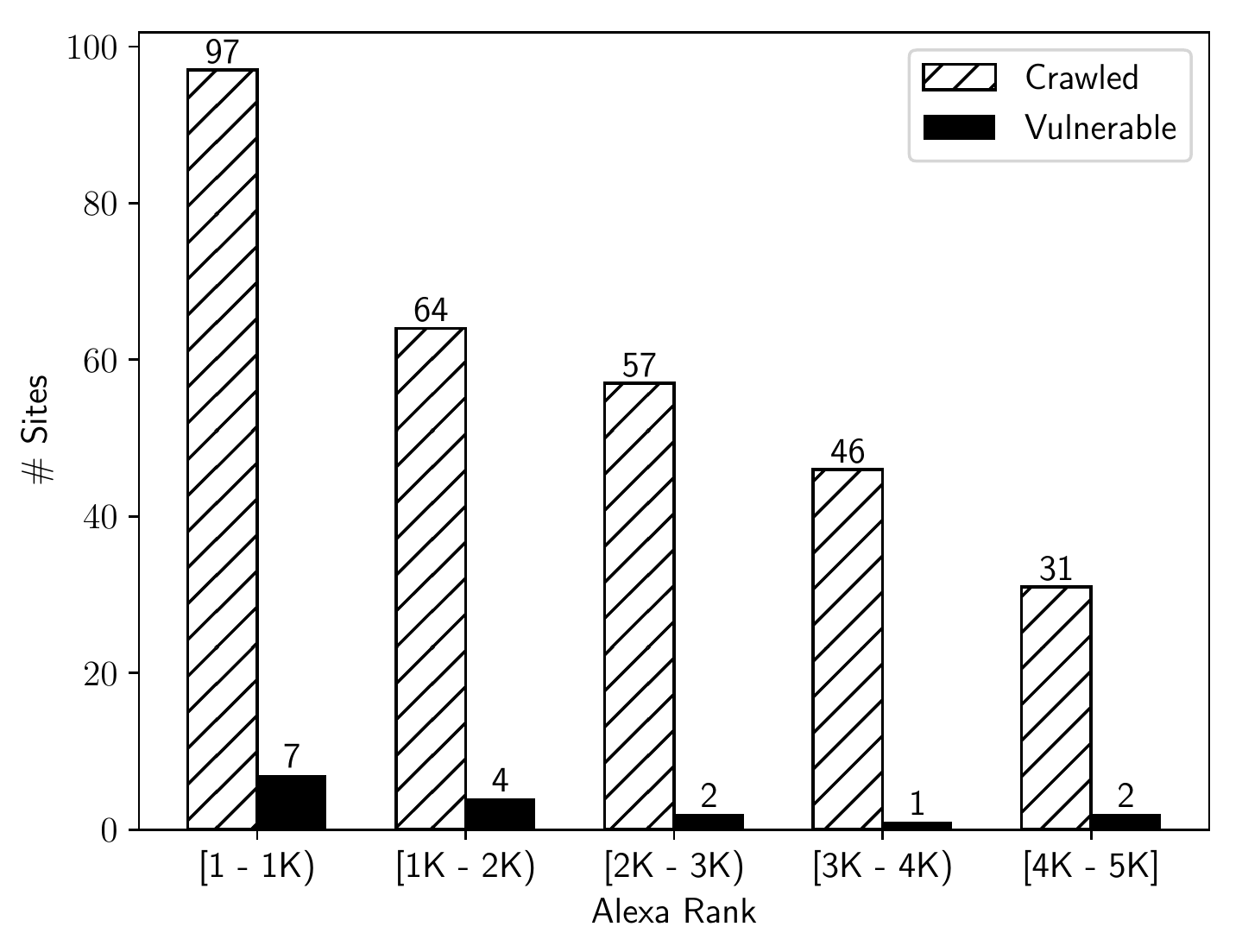}
    \caption{Distribution of the measurement data and vulnerable sites across the Alexa Top 5K.}
    \label{fig:analysis:sites}
\end{figure}

From the 295~sites comprising the collected data set, the crawler identified
16~sites (5.4\%) to contain WCD vulnerabilities. Figure~\ref{fig:analysis:sites}
presents the distribution of all sites and vulnerable sites across the Alexa Top
5K. From this, we observe that the distribution of vulnerable sites is roughly
proportional to the number of sites crawled; that is, our data does not suggest
that the incidence of WCD vulnerabilities is correlated with site popularity.

\paragraph{Content Delivery Networks (CDNs).}
\label{sec:cdn-table}
\begin{table*}[t]
    \caption{Pages, domains, and sites labeled by CDN using HTTP header heuristics. These heuristics simply check for unique vendor-specific strings added by CDN proxy servers.}
    \label{tab:analysis:cdns}
    \centering
	\footnotesize
    \setlength\tabcolsep{3pt}
    \begin{tabular}{lrrrrrr}
    \toprule

    \multirow{2}{*}{\textbf{CDN}} &
    \multicolumn{3}{c}{\textbf{Crawled}} &
    \multicolumn{3}{c}{\textbf{Vulnerable}} \\

    \cmidrule[0.5pt](lr){2-4}
    \cmidrule[0.5pt](lr){5-7}

    &
    \multicolumn{1}{c}{\textbf{Pages}} &
    \multicolumn{1}{c}{\textbf{Domains}} &
    \multicolumn{1}{c}{\textbf{Sites}} &
    \multicolumn{1}{c}{\textbf{Pages}} &
    \multicolumn{1}{c}{\textbf{Domains}} &
    \multicolumn{1}{c}{\textbf{Sites}}
    \\

    \midrule

Cloudflare & 161,140 (11.0\%) & 4,996 (4.0\%) & 143 (48.4\%) & 16,234 (93.9\%) & 72 (77.4\%) & 8 (50.0\%) \\
Akamai & 225,028 (15.3\%) & 16,473 (13.2\%) & 100 (33.9\%) & 1,059 (6.1\%) & 21 (22.6\%) & 8 (50.0\%) \\
CloudFront & 100,009 (6.8\%) & 10,107 (8.1\%) & 107 (36.3\%) & 2 (<0.1\%) & 1 (1.1\%) & 1 (6.2\%) \\
Other CDNs & 244,081 (16.6\%) & 2,456 (2.0\%) & 137 (46.4\%) & 0 (0.0\%) & 0 (0.0\%) & 0 (0.0\%) \\
\midrule
Total CDN Use & 707,210 (48.1\%) & 33,675 (27.0\%) & 244 (82.7\%) & 17,293 (100.0\%) & 93 (100.0\%) & 16 (100.0\%) \\

    \bottomrule
    \end{tabular}
\end{table*}

Using a set of heuristics that searches for well-known vendor strings in HTTP
headers, we labeled each domain and site with the corresponding
CDN. Table~\ref{tab:analysis:cdns} shows the results of this labeling. Note that
many sites use multiple CDN solutions, and therefore the sum of values in the
first four rows may exceed the totals we report in the last row.

The results show that, even though WCD attacks are equally applicable to any web
cache technology, all instances of vulnerable pages we observed are served over
a CDN. That being said, vulnerabilities are not unique to any one CDN
vendor. While this may seem to suggest that CDN use is correlated with an
increased risk of WCD, we point out that 82.7\% of sites in our experiment are
served over a CDN. A more balanced study focusing on comparing CDNs to
centralized web caches is necessary to eliminate this inherent bias in our
experiment and draw meaningful conclusions. Overall, these results indicate that
CDN deployments are prevalent among popular sites, and the resulting widespread
use of web caches may in turn lead to more opportunities for WCD attacks.

\paragraph{Response Codes.}

\begin{table}[t]
    \caption{Response codes observed in the vulnerable data set.}
    \label{tab:analysis:response_codes}
    \centering
	\footnotesize
    \begin{tabular}{lrrr}
    \toprule

    \textbf{Response Code} &
    \multicolumn{1}{c}{\textbf{Pages}} &
    \multicolumn{1}{c}{\textbf{Domains}} &
    \multicolumn{1}{c}{\textbf{Sites}}
    \\

    \midrule

    404 Not Found & 17,093 (98.8\%) & 82 (88.2\%) & 10 (62.5\%) \\
    200 Ok & 205 (1.2\%) & 19 (20.4\%) & 12 (75.0\%) \\

    \bottomrule
    \end{tabular}
\end{table}

Table~\ref{tab:analysis:response_codes} presents the distribution of HTTP
response codes observed for the vulnerable sites. This distribution is dominated
by \texttt{404 Not Found} which, while perhaps unintuitive, is indeed allowed
behavior according to RFC~7234~\cite{ietfcache}. On the other hand, while only
12~sites leaked resources with a \texttt{200 OK} response, during our manual
examination of these vulnerabilities (discussed below) we noted that more PII 
was leaked from this category of resource.

\paragraph{Cache Headers.}

\begin{table*}[t]
    \caption{Cache headers present in HTTP responses collected from vulnerable sites.}
    \label{tab:analysis:cache_headers}
    \centering
    \footnotesize
    \begin{tabular}{lrrr}
    \toprule

    \textbf{Header} &
    \multicolumn{1}{c}{\textbf{Pages}} &
    \multicolumn{1}{c}{\textbf{Domains}} &
    \multicolumn{1}{c}{\textbf{Sites}}
    \\

    \midrule

    Expires: & 1,642 (9.5\%) & 23 (24.7\%) & 13 (81.2\%) \\
    \midrule
    Pragma: no-cache & 652 (3.8\%) & 11 (11.8\%) & 6 (37.5\%) \\
    \midrule
    Cache-Control: & 1,698 (9.8\%) & 26 (28.0\%) & 14 (87.5\%) \\
    \hspace{3mm}max-age=, public & 1,093 (6.3\%) & 10 (10.8\%) & 7 (43.8\%) \\
    \hspace{3mm}max-age= & 307 (1.8\%) & 1 (1.1\%) & 1 (6.2\%) \\
    \hspace{3mm}must-revalidate, private & 102 (0.6\%) & 1 (1.1\%) & 1 (6.2\%) \\
    \hspace{3mm}max-age=, no-cache, no-store & 67 (0.4\%) & 3 (3.2\%) & 2 (12.5\%) \\
    \hspace{3mm}max-age=, no-cache & 64 (0.4\%) & 4 (4.3\%) & 1 (6.2\%) \\
    \hspace{3mm}max-age=, must-revalidate & 51 (0.3\%) & 1 (1.1\%) & 1 (6.2\%) \\
    \hspace{3mm}max-age=, must-revalidate, no-transform, private & 5 (<0.1\%) & 3 (3.2\%) & 1 (6.2\%) \\
    \hspace{3mm}no-cache & 5 (<0.1\%) & 2 (2.2\%) & 1 (6.2\%) \\
    \hspace{3mm}max-age=, private & 3 (<0.1\%) & 1 (1.1\%) & 1 (6.2\%) \\
    \hspace{3mm}must-revalidate, no-cache, no-store, post-check=, pre-check= & 1 (<0.1\%) & 1 (1.1\%) & 1 (6.2\%) \\
    \midrule
    All & 1,698 (9.8\%) & 26 (28.0\%) & 14 (87.5\%) \\
    \midrule
    (none) & 15,595 (90.2\%) & 67 (72.0\%) & 3 (18.8\%) \\

    \bottomrule
    \end{tabular}
\end{table*}

Table~\ref{tab:analysis:cache_headers} shows a breakdown of cache-relevant
headers collected from vulnerable sites. In particular, we note that despite the
presence of headers whose semantics prohibit caching---e.g., ``\texttt{Pragma:
  no-cache}'', ``\texttt{Cache-Control: no-store}''---pages carrying these
headers are cached regardless, as they were found to be vulnerable to WCD. This
finding suggests that site administrators indeed take advantage of the
configuration controls provided by web caches that allow sites to override
header-specified caching policies.

A consequence of this observation is that user-agents cannot use cache headers
to determine with certainty whether a resource has in fact been cached or not.
This has important implications for WCD detection tools that rely on cache
headers to infer the presence of WCD vulnerabilities.

\subsection{Vulnerabilities}
\label{sec:analysis:vulns}

Table~\ref{tab:analysis:results} presents a summary of the types of
vulnerabilities discovered in the collected data, labeled by manual examination.

\begin{table}[t]
    \caption{Types of vulnerabilities discovered in the data.}
    \label{tab:analysis:results}
    \centering
	\footnotesize
    \begin{tabular}{lrrrrrrrrr}
    \toprule

    \textbf{Leakage} &
    \multicolumn{1}{c}{\textbf{Pages}} &
    \multicolumn{1}{c}{\textbf{Domains}} &
    \multicolumn{1}{c}{\textbf{Sites}}
    \\

    \midrule

    PII & 17,215 (99.5\%) & 88 (94.6\%) & 14 (87.5\%) \\
    \hspace{3mm}User & 934 (5.4\%) & 17 (18.3\%) & 8 (50.0\%) \\
    \hspace{3mm}Name & 16,281 (94.1\%) & 71 (76.3\%) & 7 (43.8\%) \\
    \hspace{3mm}Email & 557 (3.2\%) & 10 (10.8\%) & 6 (37.5\%) \\
    \hspace{3mm}Phone & 102 (0.6\%) & 1 (1.1\%) & 1 (6.2\%) \\
    \midrule
    CSRF & 130 (0.8\%) & 10 (10.8\%) & 6 (37.5\%) \\
    \hspace{3mm}JS & 59 (0.3\%) & 5 (5.4\%) & 4 (25.0\%) \\
    \hspace{3mm}POST & 72 (0.4\%) & 5 (5.4\%) & 3 (18.8\%) \\
    \hspace{3mm}GET & 8 (<0.1\%) & 4 (4.3\%) & 2 (12.5\%) \\
    \midrule
    Sess.~ID / Auth.~Code & 1,461 (8.4\%) & 11 (11.8\%) & 6 (37.5\%) \\
    \hspace{3mm}JS & 1,461 (8.4\%) & 11 (11.8\%) & 6 (37.5\%) \\
    \midrule
    Total & 17,293 & 93 & 16 \\

    \bottomrule
    \end{tabular}
\end{table}

\paragraph{PII.}

14~of the 16~vulnerable sites leaked PII of various kinds, including names,
usernames, email addresses, and phone numbers. In addition to these four main
categories, a variety of other categories of PII were found to be leaked. Broad
examples of other PII include financial information (e.g., account balances,
shopping history) and health information (e.g., calories burned, number of
steps, weight). While it is tempting to dismiss such information as trivial, we
note that PII such as the above can be used as the basis for highly effective
spearphishing attacks%
~\cite{downs2006decision,jagatic2007social,hong2012state,caputo2014going}.

\paragraph{Security Tokens.}

Using the entropy-based procedure described in Section~\ref{sec:methodology}, we
also analyzed the data for the presence of leaked security tokens. Then, we
manually verified our findings by accessing the vulnerable sites using a browser
and checking for the presence of the tokens suspected to have been
leaked. Finally, we manually verified representative examples of each class of
leaked token for exploitability using the test accounts established during the
measurement.

6~of the 16~vulnerable sites leaked CSRF tokens valid for a session, which could
allow an attacker to conduct CSRF attacks despite the presence of a deployed
CSRF defense. 3~of these were discovered in hidden form elements used to protect
POST requests, while an additional 4~were found in inline JavaScript that was
mostly used to initiate HTTP requests. We also discovered 2~sites leaking CSRF
tokens in URL query parameters for GET requests, which is somewhat at odds with
the convention that GET requests should be idempotent.

6~of the 16~vulnerable sites leaked session identifiers or user-specific API
tokens in inline JavaScript. These session identifiers could be used to
impersonate victim users at the vulnerable site, while the API tokens could be
used to issue API requests as a victim user.

\paragraph{Authenticated vs.~Unauthenticated Attackers.}

The methodology we described in Section~\ref{sec:methodology} includes a detection
step intended to discover whether a suspected WCD vulnerability was exploitable
by an unauthenticated user by accessing a cached page without sending any stored
session identifiers in the requests. In only a few cases did this automated
check fail; that is, in virtually every case the discovered vulnerability was
exploitable by an unauthenticated user. Even worse, manual examination of the
failure cases revealed that in each one the crawler had produced a false
negative and that in fact all of the remaining vulnerabilities were exploitable
by unauthenticated users as well. This implies that WCD, as a class of
vulnerability, tends not to require an attacker to authenticate to a vulnerable
site in order to exploit those vulnerabilities. In other words, requiring strict
account verification through credentials such as valid SSNs or credit card
numbers is not a viable mitigation for WCD.

\subsection{Study Summary}
\label{sec:analysis:conclusions}

Summarizing the major findings of this first experiment, we found that 16 out of
295~sites drawn from the Alexa Top 5K contained web cache deception~(WCD)
vulnerabilities. We note that while this is not a large fraction of the sites
scanned, these sites have substantial user populations as to be expected with
their placement in the Alexa rankings. This, combined with the fact that WCD
vulnerabilities are relatively easy to exploit, leads us to conclude that these
vulnerabilities are serious and that this class of vulnerability deserves
attention from both site administrators and the security community.

We found that the presence of cache headers was an unreliable indicator for
whether a resource is cached, implying that existing detection tools relying on
this signal may inadvertently produce false negatives when scanning sites for
WCD vulnerabilities. We found vulnerable sites to leak PII that would be useful
for launching spearphishing attacks, or security tokens that could be used to
impersonate victim users or bypass important web security defenses. Finally, the
WCD vulnerabilities discovered here did not require attackers to authenticate to
vulnerable sites, meaning sites with restrictive sign-up procedures are not
immune to WCD vulnerabilities.

\section{Variations on Path Confusion}
\label{sec:advanced_path_confusion}

\begin{figure}[t]
\centering
{\sffamily

\begin{subfigure}[t]{1\columnwidth}
\begin{lstlisting}[basicstyle=\fontsize{8}{8}\ttfamily]
example.com/account.php
example.com/account.php**/**@@nonexistent.css@@
\end{lstlisting}
\caption{Path Parameter}
\label{fig:confusion:parameter}
\end{subfigure}

\begin{subfigure}[t]{1\columnwidth}
\begin{lstlisting}[basicstyle=\fontsize{8}{8}\ttfamily]
example.com/account.php
example.com/account.php**%0A**@@nonexistent.css@@
\end{lstlisting}
\caption{Encoded Newline (\texttt{\symbol{92}n})}
\label{fig:confusion:newline}
\end{subfigure}

\begin{subfigure}[t]{1\columnwidth}
\begin{lstlisting}[basicstyle=\fontsize{8}{8}\ttfamily]
example.com/account.php;par1;par2
example.com/account.php**%3B**@@nonexistent.css@@
\end{lstlisting}
\caption{Encoded Semicolon (\texttt{;})}
\label{fig:confusion:semicolon}
\end{subfigure}

\begin{subfigure}[t]{1\columnwidth}
\begin{lstlisting}[basicstyle=\fontsize{8}{8}\ttfamily]
example.com/account.php#summary
example.com/account.php**%23**@@nonexistent.css@@
\end{lstlisting}
\caption{Encoded Pound (\texttt{\#})}
\label{fig:confusion:pound}
\end{subfigure}

\begin{subfigure}[t]{1\columnwidth}
\begin{lstlisting}[basicstyle=\fontsize{8}{8}\ttfamily]
example.com/account.php?name=val
example.com/account.php**%3F**name=val@@nonexistent.css@@
\end{lstlisting}
\caption{Encoded Question Mark (\texttt{?})}
\label{fig:confusion:question}
\end{subfigure}

\caption{Five practical \bluebold{path confusion} techniques for crafting URLs
  that reference \redbold{nonexistent file names}. In each example, the first
  URL corresponds to the regular page, and the second one to the malicous URL
  crafted by the attacker. More generally, \redbold{nonexistent.css} corresponds
  to a nonexistent file where \redbold{nonexistent} is an arbitrary string and
  \redbold{.css} is a popular static file extension such as .css, .txt, .jpg,
  .ico, .js etc.}

\label{fig:confusion_techniques}
}
\end{figure}

Web cache technologies may be configured to make their caching decisions based
on complex rules such as pattern matches on file names, paths, and header
contents. Launching a successful WCD attack requires an attacker to craft a
malicious URL that triggers a caching rule, but also one that is interpreted as
a legitimate request by the web server. Caching rules often cannot be reliably
predicted from an attacker's external perspective, rendering the process
of crafting an attack URL educated guesswork.

Based on this observation, we hypothesize that exploring variations on the path
confusion technique may increase the likelihood of triggering caching rules and
a valid web server response, and make it possible to exploit additional WCD
vulnerabilities on sites that are not impacted by the originally proposed
attack. To test our hypothesis, we performed a second round of measurements
fourteen months after the first experiment, in July, 2019.

Specifically, we repeated our methodology, but tested payloads crafted with
different path confusion techniques in an attempt to determine how many more
pages could be exploited with path confusion variations. We used an extended
seed pool for this study, containing 295~sites from the original set and an
additional 45 randomly selected from the Alexa Top 5K, for a total of~340. In
particular, we chose these new sites among those that \textit{do not} use Google
OAuth in an attempt to mitigate potential bias in our previous measurement. One
negative consequence of this decision was that we had to perform the account
creation step entirely manually, which limited the number of sites we could
include in our study in this way. Finally, we revised the URL grouping
methodology by only selecting and exploiting a page among the first 500~pages
when there is at least one marker in the content, making it more efficient for
our purposes, and less resource-intensive on our targets. In the following, we
describe this experiment and present our findings.

\subsection{Path Confusion Techniques}

Recall from our analysis and Table~\ref{tab:analysis:response_codes} that our
WCD tests resulted in a \texttt{404 Not Found} status code in the great majority
of cases, indicating that the web server returned an error page that is less
likely to include PII. In order to increase the chances of eliciting
a \texttt{200 OK} response while still triggering a caching rule, we propose
additional path confusion techniques below based on prior
work~\cite{accunetix2019reverseproxy,tsai1,tsai2}), also illustrated in
Figure~\ref{fig:confusion_techniques}. Note that \textit{Path Parameter} in the
rest of this section refers to the original path confusion technique discussed
in this work.

\paragraph{Encoded Newline (\texttt{\symbol{92}n}).}

Web servers and proxies often (but not always) stop parsing URLs at a newline
character, discarding the rest of the URL string. For this path confusion
variation, we use an encoded newline (\texttt{\%0A}) in our malicious URL (see
Figure~\ref{fig:confusion:newline}). We craft this URL to exploit web servers that
drop path components following a newline (i.e., the server sees
\texttt{example.com/account.php}), but are fronted by caching proxies that
instead do not properly decode newlines (the proxy sees
\texttt{example.com/account.php\%0Anonexistent.css}). As a result, a request for
this URL would result in a successful response, and the cache would store the
contents believing that this is static content based on the nonexistent file's
extension.

\paragraph{Encoded Semicolon (\texttt{;}).}

Some web servers and web application frameworks accept lists of parameters in
the URL delimited by semicolons; however, the caching proxy fronting the server
may not be configured to recognize such lists. The path confusion technique we
present in Figure~\ref{fig:confusion:semicolon} exploits this scenario by
appending the nonexistent static file name after a semicolon. In a successful
attack, the server would decode the URL and return a response for
\texttt{example.com/account.php}, while the proxy would fail to decode the
semicolon, interpret \texttt{example.com/account.php\%3Bnonexistent.css} as a
resource, and attempt to cache the nonexistent style sheet.

\paragraph{Encoded Pound (\texttt{\#}).}

Web servers often process the pound character as an HTML fragment identifier,
and therefore stop parsing the URL at its first occurrence. However, proxies and
their caching rules may not be configured to decode pound signs, causing them to
process the entire URL string. The path confusion technique we present in
Figure~\ref{fig:confusion:pound} once again exploits this inconsistent
interpretation of the URL between a web server and a web cache, and works in a
similar manner to the encoded newline technique above. That is, in this case the
web server would successfully respond for \texttt{example.com/account.php},
while the proxy would attempt to cache
\texttt{example.com/account.php\%23nonexistent.css}.

\paragraph{Encoded Question Mark (\texttt{?}).}

This technique, illustrated in Figure~\ref{fig:confusion:question}, targets
proxies with caching rules that are not configured to decode and ignore standard
URL query strings that begin with a question mark. Consequently, the web server
would generate a valid response for \texttt{example.com/account.php} and the
proxy would cache it, misinterpreting the same URL as
\texttt{example/account.php\%3Fname=valnonexistent.css}.

\subsection{Results}

\begin{table}[t]
    \caption{Response codes observed with successful WCD attacks for each path
      confusion variation.}
    \label{tab:analysis:confusion_response_codes}
    \centering
	\footnotesize
    \begin{tabular}{lrrrrrr}
    \toprule

    \multirow{2}{*}{\textbf{Technique}} &
    \multicolumn{2}{c}{\textbf{Pages}} &
    \multicolumn{2}{c}{\textbf{Domains}} &
    \multicolumn{2}{c}{\textbf{Sites}}
    \\

    \cmidrule[0.5pt](lr){2-3}
    \cmidrule[0.5pt](lr){4-5}
    \cmidrule[0.5pt](lr){6-7}

    &
    \multicolumn{1}{c}{\textbf{200}} &
    \multicolumn{1}{c}{\textbf{!200}} &
    \multicolumn{1}{c}{\textbf{200}} &
    \multicolumn{1}{c}{\textbf{!200}} &
    \multicolumn{1}{c}{\textbf{200}} &
    \multicolumn{1}{c}{\textbf{!200}}
    \\

    \midrule

    Path Parameter & 3,870 & 25,932 & 31 & 93 & 13 & 7 \\
    Encoded \texttt{\symbol{92}n} & 1,653 & 24,280 & 79 & 76 & 9 & 7 \\
    Encoded \texttt{;} & 3,912 & 25,576 & 91 & 92 & 13 & 7 \\
    Encoded \texttt{\#} & 7,849 & 20,794 & 102 & 85 & 14 & 7 \\
    Encoded \texttt{?} & 11,282 & 26,092 & 122 & 86 & 17 & 8 \\
    All Encoded & 11,345 & 31,063 & 128 & 94 & 20 & 9 \\
    \midrule
    Total & 12,668 & 32,281 & 132 & 97 & 22 & 9 \\

    \bottomrule
    \end{tabular}
\end{table}

We applied our methodology to the seed pool of 340~sites, using each path
confusion variation shown in Figure~\ref{fig:confusion_techniques}. We also
performed the test with the Path Parameter technique, which was an identical
test case to our original experiment. We did this in order to identify those
pages that are not vulnerable to the original WCD technique, but only to its
variations.

We point out that the results we present in this second experiment for the Path
Parameter technique differ from our first measurement. This suggests that, in
the fourteen-month gap between the two experiments, either the site operators
fixed the issue after our notification, or that there were changes to the site
structure or caching rules that mitigated existing vulnerabilities or exposed
new vulnerable pages. In particular, we found 16~vulnerable sites in the
previous experiment and 25~in this second study, while the overlap between the
two is only~4.

Of the 25~vulnerable sites we discovered in this experiment, 20~were among the
previous set of 295~that uses Google OAuth, and 5~among the newly picked 45~that
do not. To test whether the incidence distributions of vulnerabilities among
these two sets of sites show a statistically significant difference, we applied
Pearson's \(\chi^2\) test, where vulnerability incidence is treated as the
categorical outcome variable and OAuth/non-OAuth site sets are comparison
groups. We obtained a test statistic of 1.07 and a p-value of 0.30, showing that
the outcome is independent of the comparison groups, and that incidence
distributions do not differ significantly at typically chosen significance
levels (i.e., p $>$ 0.05 ). That is, our seed pool selection did not bias our
findings.

\begin{table}[t]
    \caption{Vulnerable targets for each path confusion variation.}
    \label{tab:analysis:path_confusion_results}
    \centering
	\footnotesize
    \begin{tabular}{lrrrrrrrrr}
    \toprule

    \textbf{Technique} &
    \multicolumn{1}{c}{\textbf{Pages}} &
    \multicolumn{1}{c}{\textbf{Domains}} &
    \multicolumn{1}{c}{\textbf{Sites}}
    \\

    \midrule

    Path Parameter & 29,802 (68.9\%) & 103 (69.6\%) & 14 (56.0\%) \\
    Encoded \texttt{\symbol{92}n} & 25,933 (59.9\%) & 86 (58.1\%) & 11 (44.0\%) \\
    Encoded \texttt{;} & 29,488 (68.2\%) & 105 (70.9\%) & 14 (56.0\%) \\
    Encoded \texttt{\#} & 28,643 (66.2\%) & 109 (73.6\%) & 15 (60.0\%) \\
    Encoded \texttt{?} & 37,374 (86.4\%) & 130 (87.8\%) & 19 (76.0\%) \\
    All Encoded & 42,405 (98.0\%) & 144 (97.3\%) & 23 (92.0\%) \\
    \midrule
    Total & 43,258 (100.0\%) & 148 (100.0\%) & 25 (100.0\%) \\

    \bottomrule
    \end{tabular}
\end{table}

\begin{table*}[t]
    \caption{Number of unique pages/domains/sites exploited by each path
      confusion technique. Element $(i, j)$ indicates number of many pages
      exploitable using the technique in row $i$, whereas technique in
      column $j$ is ineffective.}
    \label{tab:analysis:path_confusion_unique_detection}
    \centering
	\footnotesize
    \begin{tabular}{lcccccc}
    \toprule

    \textbf{Technique} &
    \multicolumn{1}{c}{\textbf{Path Parameter}} &
    \multicolumn{1}{c}{\textbf{Encoded \texttt{\symbol{92}n}}} &
    \multicolumn{1}{c}{\textbf{Encoded \texttt{;}}} &
    \multicolumn{1}{c}{\textbf{Encoded \texttt{\#}}} &
    \multicolumn{1}{c}{\textbf{Encoded \texttt{?}}}

    \\

    \midrule

    Path Parameter & - & 4,390 / 26 / 7 & 1,010 / 5 / 4 & 5,691 / 11 / 3 & 5,673 / 12 / 3 \\
    Encoded \texttt{\symbol{92}n} & 521 / 9 / 4 & - & 206 / 5 / 3 & 3,676 / 5 / 3 & 3,668 / 5 / 3 \\
    Encoded \texttt{;} & 696 / 7 / 4 & 3,761 / 24 / 6 & - & 4,881 / 9 / 2 & 4,863 / 8 / 0 \\
    Encoded \texttt{\#} & 4,532 / 17 / 4 & 6,386 / 28 / 7 & 4,036 / 13 / 3 & - & 90 / 1 / 1 \\
    Encoded \texttt{?} & 13,245 / 39 / 8 & 15,109 / 49 / 11 & 12,749 / 33 / 5 & 8,821 / 22 / 5 & - \\
    \midrule
    All Encoded & 13,456 / 45 / 11 & 16,472 / 58 / 12 & 12,917 / 39 / 9 & 13,762 / 35 / 8 & 5,031 / 14 / 4 \\

    \bottomrule
    \end{tabular}
\end{table*}

\paragraph{\textbf{Response Codes.}}

We present the server response codes we observed for vulnerable pages in
Table~\ref{tab:analysis:confusion_response_codes}. Notice that there is a stark
contrast in the number of \texttt{200 OK} responses observed with some of the
new path confusion variations compared to the original. For instance, while
there were 3,870~success codes for Path Parameter, Encoded~\texttt{\#} and
Encoded~\texttt{?} resulted in~7,849 and 11,282~success responses
respectively. That is, two new path confusion techniques were indeed able to
elicit significantly higher numbers of successful server responses, which is
correlated with a higher chance of returning private user information.
The remaining two variations performed closer to the original technique.

\paragraph{\textbf{Vulnerabilities.}}

In this experiment we identified a total of 25~vulnerable
sites. Table~\ref{tab:analysis:path_confusion_results} shows a breakdown of
vulnerable pages, domains, and sites detected using different path confusion
variations.  Overall, the original path confusion technique resulted in a fairly
successful attack, exploiting~68.9\% of pages and 14~sites. Still, the new
techniques combined were able to exploit 98.0\%~of pages, and 23~out of~25
vulnerable sites, showing that they significantly increase the likelihood for
a successful attack.

We next analyze whether any path confusion technique was able to successfully
exploit pages that were not impacted by others. We present these results in
Table~\ref{tab:analysis:path_confusion_unique_detection} in a matrix form, where
each element $(i,j)$ shows how many pages/domains/sites were exploitable using
the technique in row $i$, whereas utilizing the technique listed in column $j$
was ineffective for the same pages/domains/sites.

The results in Table~\ref{tab:analysis:path_confusion_unique_detection} confirm
that each path confusion variation was able to attack a set of unique
pages/domains/sites that were not vulnerable to other techniques, attesting to
the fact that utilizing a variety of techniques increases the chances of
successful exploitation. In fact, of the 25~vulnerable sites, 11~were only
exploitable using one of the variations we presented here, but not the Path
Parameter technique.

All in all, the results we present in this section confirm our hypothesis that
launching WCD attacks with variations on path confusion, as opposed to only
using the originally proposed Path Parameter technique, results in an increased
possibility of successful exploitation. Moreover, two of the explored variations
elicit significantly more \texttt{200 OK} server responses in the process,
increasing the likelihood of the web server returning valid private information.

We stress that the experiment we present in this section is necessarily limited
in scale and scope. Still, we believe the findings sufficiently demonstrate that
WCD can be easily modified to render the attack more damaging, exploiting unique
characteristics of web servers and caching proxies in parsing URLs. An important
implication is that defending against WCD through configuration adjustments is
difficult and error prone. Attackers are likely to have the upper hand in
devising new and creative path confusion techniques that site operators may
not anticipate.

\section{Empirical Experiments}

Practical exploitation of WCD vulnerabilities depends on many factors such as
the caching technology used and caching rules configured. In this section, we
present two empirical experiments we performed to demonstrate the impact of
different cache setups on WCD, and discuss our exploration of the default
settings for popular CDN providers.

\begin{table*}[t]
    \caption{Default caching behavior for popular CDNs, and cache control
      headers honored by default to prevent caching.}
    \label{tab:cacheability}
    \centering
	\footnotesize
    \setlength\tabcolsep{6pt}
    \begin{tabular}{llccc}
    \toprule

    \multirow{3}{*}{\textbf{CDN}} &
    \multirow{3}{*}{\textbf{Default Cached Objects}} &
    \multicolumn{3}{c}{\textbf{Honored Headers}} \\

    \cmidrule[0.5pt](lr){3-5}
    
    &
    &
    \multicolumn{1}{c}{\textbf{no-store}} &
    \multicolumn{1}{c}{\textbf{no-cache}} &
    \multicolumn{1}{c}{\textbf{private}}  \\
    \midrule
    Akamai & Objects with a predefined list of static file extensions only.  & \xmark & \xmark & \xmark \\
    \addlinespace
    Cloudflare & Objects with a predefined list of static file extensions, \textbf{AND} & \cmark & \cmark & \cmark \\
    & all objects with cache control headers \texttt{public} or \texttt{max-age > 0}. & & \\
    \addlinespace
    CloudFront & All objects. & \cmark & \cmark & \cmark \\
    \addlinespace
    Fastly & All objects. & \xmark & \xmark & \cmark \\    
    \bottomrule
    \end{tabular}
\end{table*}

\subsection{Cache Location}

While centralized server-side web caches can be trivially exploited from any
location in the world, exploiting a distributed set of CDN cache servers is more
difficult. A successful WCD attack may require attackers to correctly target the
same edge server that their victim connects to, where the cached sensitive
information is stored. As extensively documented in existing WCD literature,
attackers often achieve that by connecting to the server of interest directly
using its IP address and a valid HTTP \texttt{Host} header corresponding to the
vulnerable site.

We tested the impact of this practical constraint by performing the
\textit{victim} interactions of our methodology from a machine located in
Boston, MA, US, and launching the attack from another server in Trento,
Italy. We repeated this test for each of the 25~sites confirmed to be vulnerable
in our second measurement described in Section~\ref{sec:advanced_path_confusion}.

The results showed that our attack failed for 19~sites as we predicted,
requiring tweaks to target the correct cache server. Surprisingly, the remaining
6~sites were still exploitable even though headers indicated that they were
served over CDNs (3~Akamai, 1~Cloudflare, 1~CloudFront, and 1~Fastly).

Upon closer inspection of the traffic, we found headers in our Fastly example
indicating that a cache miss was recorded in their Italy region, followed by a
retry in the Boston region that resulted in the cache hit, which led to a 
successful attack. We were not able to explore the remaining cases with the data
servers exposed to us.

Many CDN providers are known to use a tiered cache model, where content may be
available from a parent cache even when evicted from a
child~\cite{cache_hierarchy1,cache_hierarchy2}. The Fastly example above
demonstrates this situation, and is also a plausible explanation for the
remaining cases. Another possibility is that the vulnerable sites
were using a separate centralized server-side cache fronted by their CDN
provider. Unfortunately, without a clear understanding of proprietary CDN
internals and visibility into site owners' infrastructure, it is not feasible to
determine the exact cache interactions.

Our experiment confirms that cache location is a practical constraint for
a successful WCD attack where a distributed set of cache servers is involved,
but also shows that attacks are viable in certain scenarios without
necessitating additional traffic manipulation.

\subsection{Cache Expiration}

Web caches typically store objects for a short amount of time, and then evict
them once they expire. Eviction may also take place prematurely when web caches
are under heavy load. Consequently, an attacker may have a limited window of
opportunity to launch a successful WCD attack until the web cache drops the
cached sensitive information.

In order to measure the impact of cache expiration on WCD, we repeated the
\textit{attacker} interactions of our methodology with 1~hour, 6~hour, and 1~day
delays. \footnote{We only tested 19~sites out of 25, as the remaining 6~had
  fixed their vulnerabilities by the time we performed this experiment.}  We
found that 16, 10, and 9~sites were exploitable in each case, respectively.

These results demonstrate that exploitation is viable in realistic attack
scenarios, where there are delays between the victim's and attacker's
interactions with web caches. That being said, caches will eventually evict
sensitive data, meaning that attacks with shorter delays are more likely to be
successful. We also note that we performed this test with a randomly chosen
vulnerable page for each site as that was sufficient for our purposes. In
practice, different resources on a given site may have varying cache expiration
times, imposing additional constraints on what attacks are possible.

\subsection{CDN Configurations}

Although any web cache technology can be affected by WCD, we established in
Section~\ref{sec:cdn-table} that CDNs play a large role in cache use on the
Internet. Therefore, we conducted an exploratory experiment to understand the
customization features CDN vendors offer and, in particular, to observe their
default caching behavior. To that end, we created free or trial accounts with
four major CDN providers: Akamai, Cloudflare, CloudFront, and Fastly. We only
tested the basic content delivery solutions offered by each vendor and did not
enable add-on features such as web application firewalls.

We stress that major CDN providers offer rich configuration options, including
mechanisms for site owners to programmatically interact with their traffic. A
systematic and exhaustive analysis of CDN features and corresponding WCD vectors
is an extremely ambitious task beyond the scope of this paper. The results we
present in this section are only intended to give high-level insights into how
much effort must be invested in setting up a secure and safe CDN environment,
and how the defaults behave.

\paragraph{Configuration.}
All four CDN providers we experimented with offer a graphical interface and APIs
for users to set up their origin servers, apply caching rules, and configure how
HTTP headers are processed. In particular, all vendors provide ways to honor or
ignore Cache-Control headers, and users can choose whether to strip headers or
forward them downstream to clients. Users can apply caching decisions and
time-to-live values for cached objects based on expressions that match the
requested URLs.

Akamai and Fastly configurations are translated to and backed by domain-specific
configuration languages, while Cloudflare and CloudFront do not expose their
back-end to users. Fastly internally uses Varnish caches, and gives users full
control over the Varnish Configuration Language (VCL) that governs their
setup. In contrast, Akamai appears to support more powerful HTTP processing
features than Varnish, but does not expose all features to users
directly. Quoting an Akamai blog post: \textit{``Metadata [Akamai's
    configuration language] can do almost anything, good and bad, which is why
  WRITE access to metadata is restricted, and only Akamai employees can add
  metadata to a property configuration directly.''}~\cite{akamai_metadata}

In addition to static configurations, both Akamai and Cloudflare offer
mechanisms for users to write programs that execute on the edge server, and
dynamically manipulate traffic and caches~\cite{workers,edgeworkers}.

In general, while Cloudflare, CloudFront, and Fastly offer free accounts
suitable for personal use, they also have paid tiers that lift restrictions
(e.g., Cloudflare only supports 3 cache rules in the free tier) and provide
professional services support for advanced customization. Akamai strictly
operates in the business-to-business market where configuration is driven by a
professional services team, as described above.

\paragraph{Cacheability.}
Next, we tested the caching behavior of CDN providers with a default
configuration. Our observations here are limited to 200 OK responses pertaining
to WCD; for an in-depth exploration of caching decisions involving 4xx or 5xx
error responses, we refer readers to Nguyen et al.~\cite{nguyen2019}. We
summarize our observations in Table~\ref{tab:cacheability}, which lists the
conditions for caching objects in HTTP responses, and whether including the
relevant Cache-Control headers prevent caching.

These results show that both Akamai and Cloudflare rely on a predefined list of
static file extensions (e.g., .jpg, .css, .pdf, .exe) when making cacheability
decisions. While Cloudflare allows origin servers to override the decision in
both directions via Cache-Control headers, either to cache non-static files or
prevent caching static files, Akamai's default rule applies unconditionally.

CloudFront and Fastly adopt a more aggressive caching strategy: in the absence
of Cache-Control headers all objects are cached with a default time-to-live
value. Servers behind CloudFront can prevent caching via Cache-Control headers
as expected. However, Fastly only honors the \texttt{private} header value.

\subsection{Lessons Learned}
The empirical evidence we presented in this section suggests that configuring
web caches correctly is not a trivial task. Moreover, the complexity of
detecting and fixing a WCD vulnerability is disproportionately high compared to
launching an attack.

As we have seen above, many major CDN vendors do not make RFC-compliant caching
decisions in their default configurations~\cite{ietfcache}. Even the more
restrictive default caching rules based on file extensions are prone to security
problems; for example, both Akamai and Cloudflare could cache dynamically
generated PDF files containing tax statements if configured incorrectly. On the
other hand, we do not believe that these observations implicate CDN vendors in
any way, but instead emphasize that CDNs are not intended to be plug \& play
solutions for business applications handling sensitive data. All CDNs provide
fine-grained mechanisms for caching and traffic manipulation, and site owners
must carefully configure and test these services to meet their needs.

We reiterate that, while CDNs may be a prominent component of the Internet
infrastructure, WCD attacks impact all web cache technologies. The complexity of
configuring CDNs correctly, the possibility of multi-CDN arrangements, and other
centralized caches that may be involved all imply that defending against WCD
requires site owners to adopt a holistic view of their environment. Traditional
security practices such as asset, configuration, and vulnerability management
must be adapted to take into consideration the entire communication
infrastructure as a system.

From an external security researcher's perspective the challenge is even
greater. As we have also discussed in the cache location and expiration
experiments, reasoning about a web cache system's internals in a black box
fashion is a challenging task, which in turn makes it difficult to pinpoint
issues before they can be exploited. In contrast, attackers are largely immune
to this complexity; they often do not need to disentangle the cache structure
for a successful attack. Developing techniques and tools for reliable detection
of WCD---and similar web cache attacks---is an open research problem. We
believe a combination of systems security and safety approaches would be a
promising research direction, which we discuss next as we conclude this paper.

\section{Discussion \& Conclusion}
\label{sec:conclusion}

In this paper, we presented the first large-scale investigation of WCD
vulnerabilities in the wild, and showed that many sites among
the Alexa Top 5K are impacted. We demonstrated that the vulnerable
sites not only leak user PII but also secrets that, once stolen by an
attacker, can be used to bypass existing authentication and authorization
mechanisms to enable even more damaging web application attack scenarios.

Alarmingly, despite the severity of the potential damage, these vulnerabilities
still persist more than two years after the public introduction of the attack in
February 2017. Similarly, our second experiment showed that in the fourteen
months between our two measurements, only 12~out of 16~sites were able to
mitigate their WCD vulnerabilities, while the total number of vulnerabilities
rose to 25.

One reason for this slow adoption of necessary mitigations could be a lack of
user awareness.  However, the attention WCD garnered from security news outlets,
research communities, official web cache vendor press releases, and even
mainstream media also suggests that there may be other contributing
factors. In fact, it is interesting to note that there exists no technology or
tool proposed to date that allows site operators to reliably determine if any
part of their online architecture is vulnerable to WCD, or to close their
security gaps.  Similarly, there does not exist a mechanism for end-users and
web browsers to detect a WCD attack and protect themselves. Instead,
countermeasures are largely limited to general guidance by web cache vendors
and CDN providers for their users to configure their services in consideration
of WCD vectors, and the tools available offer limited manual
penetration-testing capabilities for site operators with domain-specific
knowledge.

We assert that the above is a direct and natural consequence of the fact that
WCD vulnerabilities are a \textit{system safety} problem. In an environment with
WCD vulnerabilities, there are no isolated faulty components; that is, web
servers, load balancers, proxies, and caches all individually perform the
functionality they are designed for.  Similarly, determining whether there is
human error involved and, if so, identifying where that lies are both
non-trivial tasks. In fact, site operators often have legitimate needs to
configure their systems in seemingly hazardous ways. For example, a global
corporation operating hundreds to thousands of machines may find it technically
or commercially infeasible to revise the Cache-Control header settings of their
individual web servers, and may be forced to instruct their CDN provider to
perform caching based purely on file names.

These are all strong indicators that the growing ecosystem of web caches, in
particular CDN-fronted web applications, and more generally highly-distributed
Internet-based architectures, should be analyzed in a manner that captures their
security and safety properties as a system. As aforementioned, venerable yet
still widely-used \textit{root cause analysis} techniques are likely to fall
short in these efforts, because there is no individual system component to blame
for the failure. Instead, security researchers should adopt a systems-centric
security analysis, examining not only individual system components but also
their interactions, expected outcomes, hazardous states, and accidents that may
result. Modeling and analyzing WCD attacks in this way, drawing from the rich
safety engineering literature~\cite{safety} is a promising future research
direction that will help the security community understand and address similar
systems-level attacks effectively.

\section*{Acknowledgments}
We thank our shepherd Ben Stock and the anonymous reviewers; this paper is all
the better for their helpful feedback. This work was supported by the National
Science Foundation under grant CNS-1703454, Secure Business Austria, ONR project
``In-Situ Malware Containment and Deception through Dynamic In-Process
Virtualization,'' and EU H2020-SU-ICT-03-2018 Project No. 830929
CyberSec4Europe.

\bibliographystyle{plain}
\bibliography{paper}

\end{document}